\documentclass{article}

\usepackage{arxiv}

\usepackage[utf8]{inputenc} 
\usepackage[T1]{fontenc}    
\usepackage{url}
\usepackage{booktabs}
\usepackage{amsfonts}
\usepackage{nicefrac}
\usepackage{microtype}
\usepackage{lipsum}
\usepackage{graphicx}
\usepackage{biblatex} 
\usepackage{float}
\usepackage{multirow}
\graphicspath{ {./images/} } 
\title{AI-Facilitated Episodic Future Thinking For Adults with Obesity}

\author{
    Sareh Ahmadi \\
    Department of Computer Science\\
    Virginia Tech\\
    Blacksburg, VA 24061\\
    \texttt{saraahmadi@vt.edu}
    \and
    Michelle Rockwell\\
    Fralin Biomedical Research Institute at VTC\\
    Roanoke, Virginia, USA\\
    \texttt{mrockwell@carilionclinic.org}
    \and
    Megan Stuart\\
    Fralin Biomedical Research Institute at VTC\\
    Virginia Tech, Roanoke, USA\\
    \texttt{stuartma@vtc.vt.edu}
    \and
    Nicki Rohani\\
  Fralin Biomedical Research Institute at VTC, Virginia Tech\\
  Roanoke, VA, USA \\
\texttt{nickirohani@vt.edu}
    \and
    Allison Tegge\\
    Fralin Biomedical Research Institute at VTC\\
    Roanoke, Virginia, USA\\
    \texttt{ategge@vt.edu}
    \and
    Xuan Wang\\
    Department of Computer Science\\
    Virginia Tech, Blacksburg, USA\\
    \texttt{xuanw@vt.edu}
    \and
    Jeffrey Stein\\
    Fralin Biomedical Research Institute at VTC\\
    Roanoke, Virginia, USA\\
    \texttt{jstein1@vtc.vt.edu}
    \and
    Edward A. Fox \\
    Department of Computer Science\\
    Virginia Tech\\
    Blacksburg, VA 24061\\
    \texttt{fox@vt.edu}
}

\begin{document}
\maketitle
\begin{abstract}
Episodic Future Thinking (EFT)
involves vividly imagining personal future events and experiences in detail. It has shown promise as an intervention to reduce delay discounting—the tendency to devalue delayed rewards in favor of immediate gratification—and to promote behavior change in a range of maladaptive health behaviors.
We present EFTeacher, an AI chatbot powered by the GPT-4-Turbo large language model, designed to generate EFT cues for users with lifestyle-related conditions. To evaluate the feasibility and usability of EFTeacher, we conducted a mixed-methods study that included usability assessments, user evaluations based on content characteristics questionnaires, and semi-structured interviews.
Qualitative findings indicate that participants perceived EFTeacher as communicative and 
supportive through an
engaging dialogue. The chatbot facilitated imaginative thinking and reflection on future goals. Participants appreciated its adaptability and personalization features, though some noted challenges such as repetitive dialogue and verbose responses. Our findings underscore the potential of large language model-based chatbots in EFT interventions targeting maladaptive health behaviors.

\end{abstract}

\section{Introduction}

Approximately 40\% of adults in the United States are currently living with obesity,  with projections suggesting that this figure will reach 50\% by 2030 \cite{lieberman2022state, revels2017predicting}. Although some progress has been made in addressing obesity, many communities still lack access to prevention and treatment resources, worsening health disparities related to obesity \cite{washington2023disparities}. Losing weight successfully depends on making choices today that are influenced by long-term goals, as most of the health benefits from weight loss come after months or even years of maintaining consistent healthy habits.

Episodic future thinking (EFT) involves vividly imagining personal future events and experiences in detail \cite{atance2001episodic}. EFT has been increasingly used as an intervention for behavior change regarding various maladaptive health behaviors, and for populations with lifestyle-related conditions, to reduce delay discounting \cite{brown2022putting}. Delay discounting (DD) refers to the tendency to choose immediate reward over greater, future benefits \cite{bickel2012excessive}. EFT interventions usually involve participants identifying several positive future events they look forward to and plan to attend at different times. They write or describe these events in detail and are encouraged to vividly imagine them when making decisions involving future trade-offs \cite{bickel2017toward}. Personalized positive future cues have been shown to significantly impact decision-making by encouraging individuals to consider longer-term consequences. These cues help expand the ``temporal window,'' which refers to the ability to think further into the future when evaluating choices \cite{bulley2017influence,chiou2017episodic, kaplan2016effects}.

Research indicates that obesity is linked to high levels of DD. For instance, high levels of DD are both currently and over time linked to having a higher body mass index (BMI), consuming more calories, and engaging in less physical activity \cite{bickel2021temporal,appelhans2019delay, garza2016impulsivity}.
Additionally, experimental studies demonstrate that interventions aimed at altering DD lead to related improvements in dietary habits, further supporting the idea of a functional connection between the two \cite{garza2016impulsivity,hollis2019episodic, snider2020reinforcer, hollis2020ecological}.

Laboratory studies have shown that engaging in EFT for a short time can effectively reduce DD in adults with obesity \cite{daniel2013future,brown2023episodic,stein2021bleak,bickel2020does,stein2017think}, and a range of other clinical disorders \cite{rosch2022promoting} . This means EFT helps people prioritize long-term goals over immediate rewards. Studies also demonstrate that brief EFT sessions can decrease the purchase and consumption of unhealthy, high-calorie foods both in controlled lab settings and in real-life situations, such as grocery shopping or dining in public \cite{hollis2019episodic,hollis2020mothers,hollis2020ecological, o2016episodic}.
Building on these findings, EFT has been adapted into a remotely delivered intervention as part of clinical trials for managing type 2 diabetes. This intervention aims to help individuals adhere to lifestyle changes recommended for treating obesity and its comorbidities \cite{sze2015web,epstein2022effects,brown2022putting}. EFT works by helping patients stay focused on their long-term health goals while reducing the appeal of immediate temptations like unhealthy foods or inactivity, which could derail their progress.

There are two common approaches to creating EFT cues for studies on delay discounting and health behaviors: interview-guided and survey-guided methods \cite{daniel2013future, sze2017bleak}. Both methods share some similarities in how the cues are generated. For instance, researchers may give specific instructions based on the study's goals and participants' behavior \cite{brown2022putting}. In clinical practice, having clinicians conduct interviews to guide patients through EFT is costly, requires significant training to ensure treatment fidelity, and limits when and where patients can access the intervention. While self-administered, survey-based approaches address these challenges by reducing costs and increasing accessibility, they introduce new concerns about maintaining consistency and effectiveness in treatment (e.g., due to terse or off-topic cues). 
Although the survey method was designed as a more scalable alternative to the human interview, poor quality cues in this self-administered method (e.g., vague or non-episodic) may reduce treatment fidelity and thus efficacy of the intervention. Although EFT is a promising behavioral intervention to facilitate weight loss and dietary improvement \cite{daniel2013future,brown2023episodic,stein2021bleak,bickel2020does,stein2017think}, significant effects on health behaviors are not always observed \cite{ bickel2020does,stein2021bleak}. Thus, research is needed to examine whether the method used to generate cues moderates these behavioral effects and user perceptions.

To address these issues, this study aims to evaluate an artificial intelligence (AI)-based method for delivering EFT.  This approach would improve consistency, scalability, accessibility, and effectiveness, allowing EFT to be implemented on a much larger scale while minimizing costs and other barriers. We aim to enhance and evaluate
an existing 
prototype
AI-based chatbot \cite{ahmadi2024ai}, referred to as EFTeacher, for EFT intervention. EFTeacher is a chatbot built on GPT-4-turbo,
a Large Language Model (LLM). EFTeacher is designed to assist participants in identifying personally meaningful future events, and in creating for them detailed text descriptions, or cues, for those events.

In this study, we seek to answer the following research questions:

\begin{itemize}
    \item RQ1: How do users evaluate the usability of EFTeacher?
    \item RQ2: What are users’ impressions of the EFT cue texts generated by EFTeacher?
    \item RQ3: How do users perceive working with an interactive AI chatbot for EFT generation? 
\end{itemize}

To address RQ1 and RQ2, we used the System Usability Scale (SUS) \cite{brooke1996sus}, recognizing that usability is a necessary foundation for any tool to be effective, including AI systems for EFT interventions. For RQ2, we analyzed specific content characteristics of the generated EFT cues generated by EFTeacher \cite{rung2020translating}, as these measures have been used in previous studies \cite{rung2020translating} to assess the effects of EFT cue texts. We address RQ3 using thematic analysis \cite{braun2006using, bowman2023using},  a widely used qualitative method for identifying and analyzing patterns in user perceptions and experiences.

To the best of our knowledge, this is the first study to apply LLMs for EFT cue generation—a key procedural component in behavioral interventions targeting delay discounting. Prior work has not developed, evaluated, or validated an AI-based method for eliciting EFT cues. This proof-of-concept study aims to evaluate the usability and acceptability of this novel approach, in line with prior work emphasizing the importance of user-centered and theory-based approaches to effectively develop AI-guided systems \cite{nadarzynski2019acceptability}.

\section{Related Work}

Conversational agents have emerged as valuable tools for delivering cost-effective and practical behavioral interventions. Studies demonstrate their effectiveness in promoting physical activity, encouraging healthy diets \cite{zhang2020artificial, oh2021systematic}, and supporting weight loss efforts \cite{chew2022use}. The design of engaging and impactful chatbot interventions remains a crucial focus for maximizing their potential \cite{cole2019understanding}. Furthermore, AI-driven conversational agents have been used in behavior change systems to create more interactive and engaging interventions \cite{li2023exploring, kumar2023exploring, jo2023understanding, nepal2024contextual}.

Human-computer interaction
 (HCI) technologies have shown promise in fostering mindfulness and enhancing well-being \cite{terzimehic2019review}. Studies highlight their ability to increase relaxation, focus, and self-awareness, thereby improving decision-making and motivating behavior change \cite{terzimehic2019review}. Mindful interaction with technology emphasizes the importance of reflection and deliberate choices during digital interactions, further enhancing personal growth and well-being. As a result, researchers have explored various approaches to designing conversational agents tailored to specific behavioral goals.

\textbf{Physical Activity and Healthy Eating:}
Olafsson et al. \cite{olafsson2020motivating} compared the impact of humorous versus non-humorous virtual agents in promoting exercise and healthy eating, finding humor can enhance motivation and engagement significantly. Similarly, Sun et al. \cite{sun2024can} investigated the impact of humor in chatbot-delivered interventions designed to increase physical activity. The study focused on user interactions with a chatbot, analyzing levels of participation, motivation, and physical activity. The findings showed that the use of humor in the chatbot helped users become more physically active. Paay et al. \cite{paay2022can} highlighted the role of digital personal assistants (DPAs) in motivating physical activity such as exercise, with virtual rewards being particularly effective.

Kocielnik et al. \cite{kocielnik2018reflection} introduced a reflection tool called the ``Reflection Companion,'' which encourages users to notice patterns in their physical activity and plan future actions. The system uses short daily conversations and visual data representations to help users recognize their habits and strategize for improvement, underscoring the value of reflective practices in promoting healthy behaviors.

\textbf{Smoking Cessation:}
Alphonse et al. \cite{alphonse2022exploring} revealed that human-like traits in chatbots promote accountability and engagement, making them effective tools for smoking cessation. Their qualitative study explored smokers’ experiences with Quit Coach, a quick-response chatbot integrated into a smoking cessation app. The findings suggest that the chatbot’s design increased users’ sense of accountability and engagement, supporting the use of chatbots in smoking cessation interventions.
Kjaerulff et al. \cite{kjaerulff2024exploring} explored voice-based user interfaces (VUIs) that facilitated mindfulness techniques for smoking cessation. Participants used the VUI to track cravings, engage in mindfulness exercises, or reflect on smoking habits by listening to pre-recorded audio responses. Results indicated that both VUIs and  augmented VUIS (AVUIs) show promise in supporting smoking cessation and proposed integrating Large Language Models (LLMs) for enhanced personalization and adaptability.

\textbf{Applications in Chronic Disease Management:}
Conversational agents have also shown promise in managing chronic conditions like diabetes. Bin et al. \cite{bin2023towards} proposed a chatbot design framework that addresses limitations in existing systems. This framework was developed using surveys, interviews, and a review of current healthcare chatbots. By incorporating insights from the interviews, the framework was designed to meet user needs and preferences, thereby improving the chatbot’s effectiveness in preventing type 2 diabetes. Baptista \cite{baptista2020acceptability} evaluated the acceptability of an embodied conversational agent (ECA) for type 2 diabetes self-management. Using both survey data and participant interviews, the study found that users appreciated the agent's friendly and motivational support, though improvements in communication and appearance were recommended, highlighting the potential of ECAs in chronic disease management.

\textbf{Supporting Well-Being and Mindfulness in Food-Related Activities:}
Studies have also examined how conversational agents can enhance well-being during food-related activities. Parra et al. \cite{parra2023enhancing} explored the potential of agents to promote mindfulness during meals, especially following the impact of COVID-19 on eating habits. The study began by examining how the COVID-19 pandemic affected young adults’ eating habits, noting its negative impact on well-being. Based on these findings, the researchers developed two interactive systems. Both systems demonstrated that conversational agents could make these activities more enjoyable and foster mindfulness. The researchers aim to develop more advanced agents using LLMs
and activity recognition to adapt to users’ needs effectively. Zhang et al.  \cite{zhang2023facilitating} presented a pilot study investigating how voice assistants can encourage mindful eating. Researchers created two versions of a voice assistant app: one with a friendly approach and another with a counseling approach. They tested these apps to evaluate user preferences and how the assistant's social role influenced their mindful eating experience. The study found that participants preferred the ``friend'' version, offering insights for designing voice assistants that support mindfulness practices.

\textbf{Advancing Conversational Agents for Personal Growth:}
Several studies focus on refining conversational agents to support broader personal development goals.
Saravanan et al. \cite{saravanan2022giving} explored the design and evaluation of a conversational system for a social robot employing motivational interviewing to encourage healthy eating. Their findings demonstrated that referencing past interactions in motivational interviewing sessions improved user engagement and behavior change. Participants interacted with the virtual robot over three sessions, during which their motivation levels and goal progress were measured. Results indicated that varied references to previous sessions significantly increased motivation for behavior change.
Meyer et al. \cite{meyer2023towards} highlighted the potential of conversational agents trained to recognize behavior-change language using motivational interview concepts.
Browne et al. \cite{browne2024reflective} developed and tested a conversational AI health coaching system integrated into an app. The study evaluated the system’s performance, user interaction, and the effectiveness of prompt engineering techniques. 
They
assessed the app's impact on mental well-being, finding improvements in positive affect, reduced negative affect, and enhanced self-reflection. The study concludes with a discussion of the app's usability, user feedback, and suggestions for future improvements. 

\textbf{Large Language Models in Conversational Systems:}
Recent advancements in
LLMs have further expanded the capabilities of conversational agents. Arakawa et al. \cite{arakawa2024coaching} explored LLM-powered chatbots in executive coaching, emphasizing their accessibility and logical reasoning while highlighting the importance of human-AI collaboration to promote deep self-reflection.
James et al. \cite{james2024toward} evaluated LLMs for generating personalized goals in gamified mHealth apps. A study comparing LLM-generated and pre-determined goals highlighted the scalability and efficiency of LLM-driven content creation. Nepal et al. \cite{nepal2024mindscape} introduced the MindScape app, which uses AI-generated journaling prompts tailored to smartphone behavioral data. The study showed improvements in mental well-being and self-reflection, emphasizing the app’s potential for enhancing traditional journaling for effective interventions that promote self-awareness and holistic well-being.

\section{Methodology}
\begin{figure}[h]
  \centering
  \includegraphics[width=0.7\linewidth]{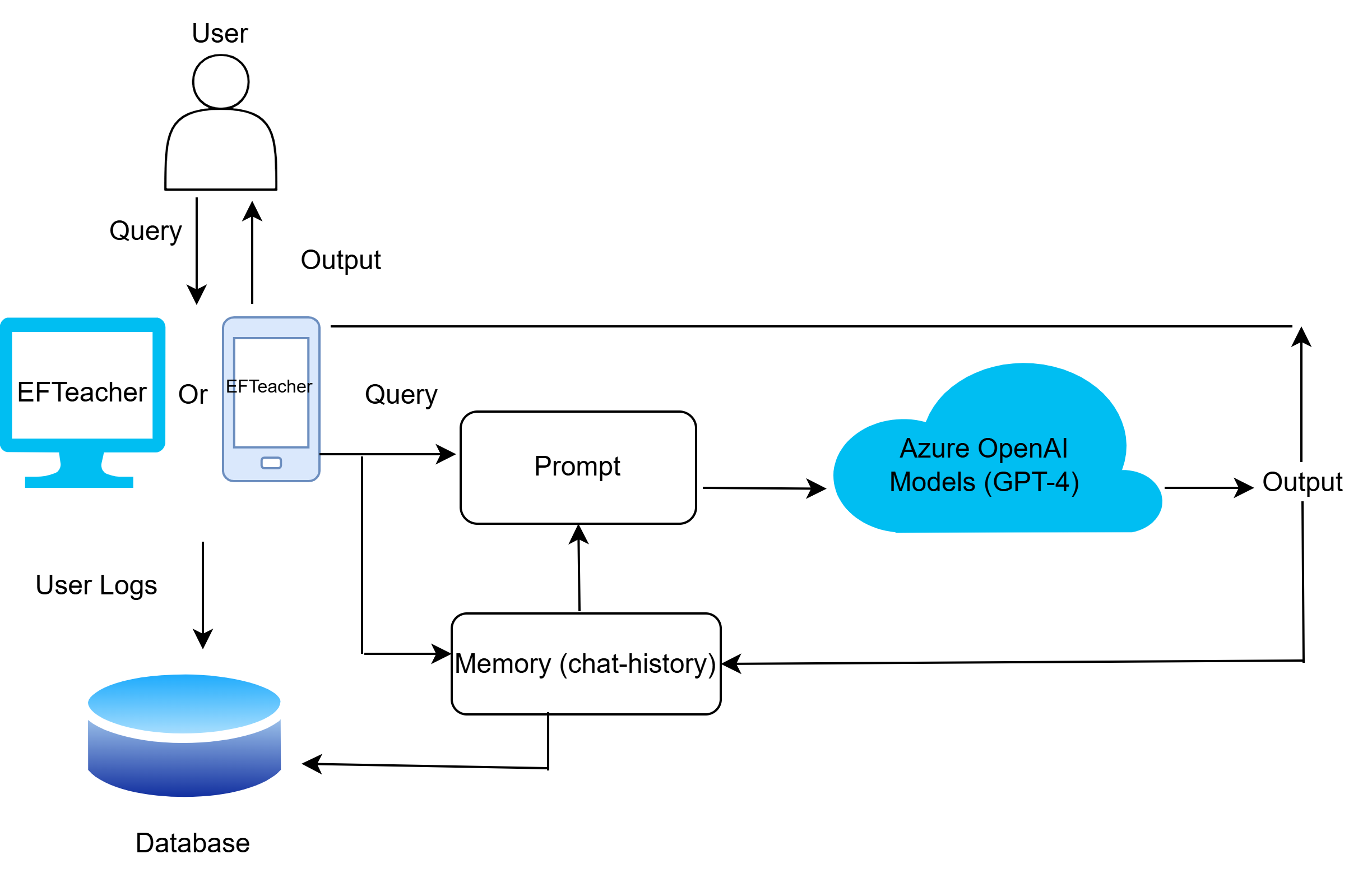}
  \caption{User Study System}

\label{system}
\end{figure}

An AI chatbot was developed to generate EFT cues \cite{ahmadi2024ai}; EFTeacher
is modeled after the clinician-led EFT interviews \cite{daniel2013future,hollis2020ecological, bickel2020does} and self-administered surveys \cite{stein2021bleak, brown2023episodic, ruhi2022episodic} used previously. We implemented a website 
for a single-session study in which participants were given a username and password and could log into the website as shown in Figure \ref{system}
to complete the AI-assisted EFT interview.
We used GPT-4-Turbo \cite{achiam2023gpt} through Microsoft Azure, which is a university-approved cloud service, for EFTteacher. 
As shown in Figure \ref{system}, memory, in the form of a list, is the history of the conversation which is passed into the prompt along with the user query (message) so the chatbot can generate an output. As depicted in the diagram, the memory stores both the
responses generated by the chatbot and the corresponding user queries. 
Each interaction between the human and GPT-4 is added
to this list so that it can be used in the next input for the model.

\begin{figure}[h]
  \centering
  \includegraphics[width=\linewidth]{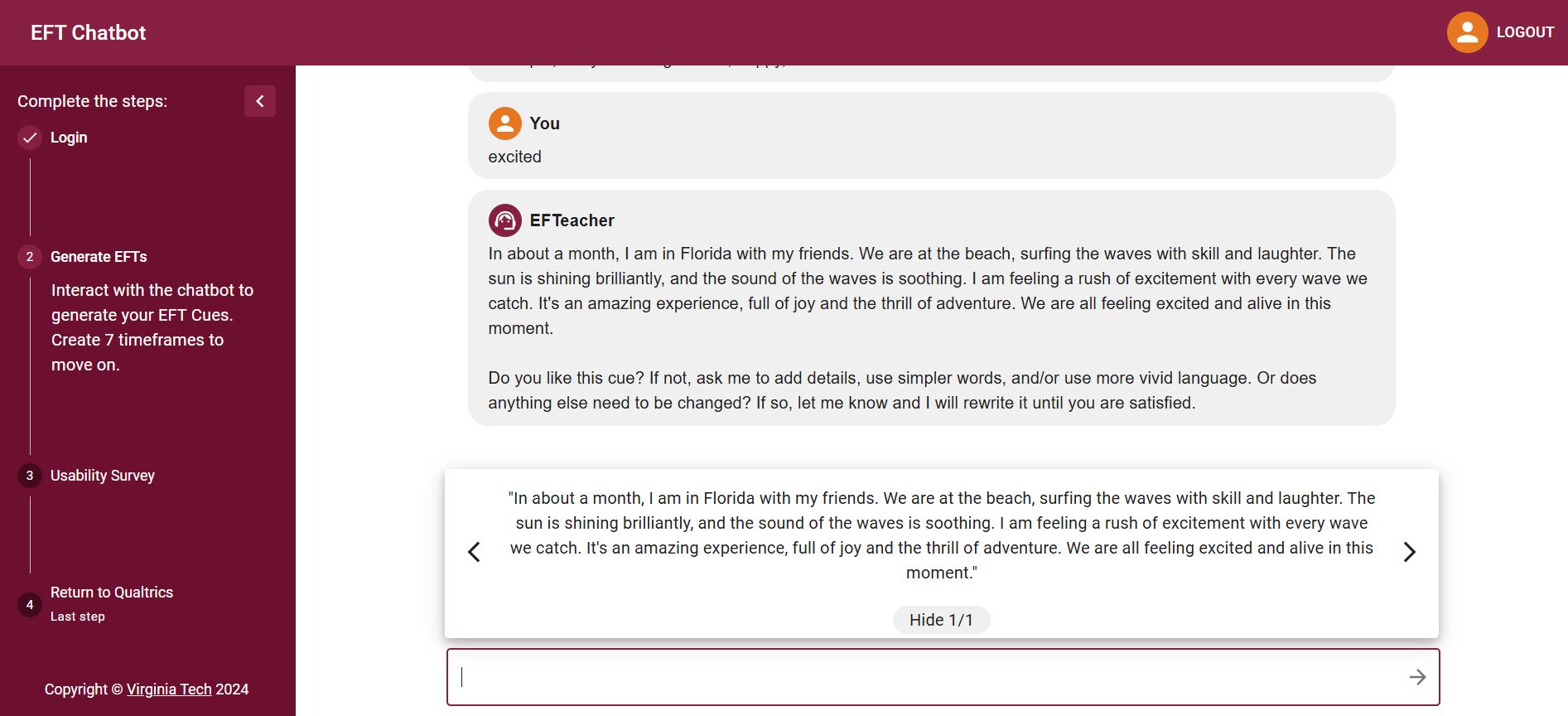}
  \caption{Chatbot Interaction Page}
  \label{wireframe}
\end{figure}

On the chat page of the website, depicted in Figure \ref{wireframe}, users interacted with the chatbot, which guided them in generating a set of EFT cue texts. 
EFTeacher first introduced EFT cues to participants.  They were then provided with clear criteria for crafting a good EFT cue. Participants were guided to ensure their cues met specific criteria: the event should be something they look forward to, vividly imaginable, confined to a single 24-hour period, and written in the present tense as if they are currently experiencing it.

To help them understand, participants were shown examples of well-crafted EFT statements that met all the criteria, as well as poor examples. They were then prompted to create their own EFT cues for specific timeframes, starting from one month into the future, ensuring their events were realistic, specific, and followed the given guidelines. This step-by-step approach ensured participants understood EFT cues and could create effective ones.

For each time-frame, participants first provided brief contextual statements about the event. Then they answered a series of standardized questions to add specific details, such as where the event was taking place, who was present, what actions were occurring, and how they felt in that moment.  
EFTeacher then combined these brief statements and detailed responses into a coherent cue, typically 3-4 sentences long. If participants gave incomplete or unrelated answers, EFTeacher prompted them to revise their responses to ensure the quality and relevance of the cues. Once an EFT cue was generated, EFTeacher sought feedback from the participant to adapt the cue based on their preferences, ensuring it aligned with their personal perspective and needs.
The user and chatbot interacted iteratively until a desired EFT cue was generated and accepted by the user. Once finalized, the EFT cue was extracted through a predefined routine and displayed in a dedicated box (cue box), as shown in Figure \ref{wireframe}.
As the conversation progressed and an EFT cue was generated for any time-frame, users could view the cues in the cue box,  navigate back to previous cues, and review them. This allowed users to keep track of which events had already been discussed during the EFT generation dialog.
For this study, once users completed covering events for seven different time-frames (one month, three months, one year, three years, five years, and ten years), the conversation concluded.

Upon completing the conversation, the chatbot's logs, including the entire conversation (referred to as memory), were saved as a file in the database. A designated routine stored the extracted EFT cues in the database, and then sent them to Qualtrics.
There
they evaluated the system's usability using System Usability Scale (SUS) \cite{brooke1996sus} questions.
Also in
Qualtrics, participants were presented with the EFT cue texts generated by EFTeacher, organized by the different time-frames. For each cue, participants were asked to evaluate its characteristics on a 5-point Likert scale, ranging from 1 (not at all) to 5 (extremely). They rated each cue based on four key aspects: Liking, Importance, Excitement, and Vividness. Afterward, the participants were presented with a survey to provide additional feedback about their experience by answering open-ended questions, as explained in detail in Section \ref{dataCollection}.

\subsection{Participants}
\begin{table}[t]
  \centering
  \caption{Summary of Participant Demographics}
  \label{study1participantdemographics}
  \begin{tabular}{lp{3.5cm} lp{3.5cm}}
    \toprule
    \textbf{Characteristic} & \textbf{Complete (n=25)} & \textbf{Characteristic} & \textbf{Complete (n=25)} \\
    \midrule
    \textbf{BMI (kg/m\textsuperscript{2}), mean (SD)} & 38.8 (8.1) & \textbf{Education, n (\%)} & \\
    \textbf{Age, mean (SD)} & 38.5 (10.1) & HS/GED or less & 7 (30.1) \\
    \textbf{Gender, n (\%)} & & Some college & 6 (23.1) \\
    Male & 10 (38.5) & 2-yr college degree & 3 (11.5) \\
    Female & 15 (61.5) & 4-yr college degree & 7 (26.9) \\
    \textbf{Household income, n (\%)} & & Grad/professional degree & 2 (7.7) \\
    $<$50,000 & 7 (30.1) & \textbf{Race, n (\%)} & \\
    \$50,000--\$79,999 & 6 (23.1) & White & 17 (65.4) \\
    \$80,000--\$119,999 & 7 (27.0) & Black/AA & 3 (15.4) \\
    $\geq$120,000 & 5 (19.2) & Asian & 1 (3.8) \\
    \textbf{Household size, median (IQR)} & 2 (2, 4) & More than one race & 4 (15.4) \\
    \textbf{Ethnicity, n (\%)} & Hispanic/Latino & 2 (7.7) \\
    \textbf{Glycemic status, n (\%)} & & \\
    Type 2 diabetes & 11 (42.3) & & \\
    Prediabetes & 6 (23.1) & & \\
    Normoglycemia & 8 (34.6) & & \\
    \bottomrule
  \end{tabular}
\end{table}

Following approval from the Institutional Review Board (IRB),
and extensive pilot testing,
 the EFT research team carried out an evaluation study with 25 participants, specifically targeting individuals with obesity. Eligibility was assessed based on the following inclusion criteria: a BMI (measured in kg/m) of 30 or higher, fluency in English, and an age range of 18 to 65 years. Participants were invited by email from a database of individuals who had previously completed eligibility questionnaires for
 our many prior studies. All participants had provided consent to be contacted about research studies. The recruitment process began with 346 email invitations  of which 17.9\% (62 individuals) started the screening questionnaire—to evaluate eligibility. Out of those, 15.3\% (53 individuals) completed the questionnaire, resulting in 9.8\% (34 individuals) being deemed eligible and providing consent to participate. Ultimately, 7.2\% (25 individuals) completed the study.  The demographics of the participants are detailed in Table \ref{study1participantdemographics}.
 Household size is summarized using the median and interquartile range (IQR), where the median represents the middle value and the IQR (25th–75th percentile) shows the range in which the central 50\% of the data falls. In general, the set of participants seemed to be relatively balanced across key demographic characteristics, in part because the recruitment proceeded in small steps, with a shifting focus to help achieve balance.

 Users began by completing an informed consent process on the Qualtrics platform \cite{qualtrics}
 where they provided their demographic information. They were then redirected to our website, where each user was assigned a unique ID stored in the database for use during their interactions.

\subsection{Data Collection}\label{dataCollection}

In this study, one of the goals was to better understand people's experiences with  EFTeacher, particularly their interactions with it. To do this, we used thematic analysis to perform a qualitative analysis, identifying themes from open-ended survey responses. 

Participants were asked five open-ended questions to share their thoughts and feelings in their own words about using EFTeacher.
To add context to these responses, participants first rated their agreement with several statements on a scale from ``Strongly Disagree'' to ``Strongly Agree.'' These statements focused on how EFTeacher's responses influenced their ability to imagine and describe future events. For example, prompts included questions like whether EFTeacher's responses felt human-like, helped them imagine future events, or made those events more vivid.

After these ratings, participants were asked to elaborate on why they felt the way they did. They were also asked to identify features of EFTeacher they found helpful and engaging, as well as those they found frustrating or less useful. This included listing at least two features they liked and believed should remain in future versions of the chatbot and two features they disliked and thought should be improved.
Finally, participants could share any additional feedback or thoughts about their overall experience with EFTeacher. This process aimed to gather detailed insights into the user experience, including both positive and negative aspects. This feedback would help refine EFTeacher and inform the design and evaluation of similar chatbots in the future.
Each participant was assigned an ID number from 100-150.

Data was also collected
for both SUS and content characteristics. 
For the SUS \cite{brooke1996sus}, we calculated the mean and standard deviation (SD) across all 25 users. For key content characteristics (i.e., Vividness, Liking, Excitement, and Importance), we calculated the median of these four items from each EFT cue that was generated for a specific timeframe, where users rated each item on a scale of 1 (not at all) to 5 (extremely) based on how they felt about each cue text.

\subsection{Users' Usability  Experience}
\begin{figure}[h]
  \centering
  \includegraphics[scale=0.4]{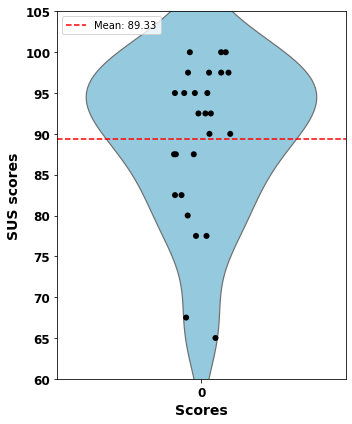} 
  \caption{SUS Scores}
  \label{sus}
\end{figure}

To address
RQ1 (i.e., 
How do users evaluate the usability  of EFTeacher?),
the usability of the chatbot was assessed using 10 standard SUS items on a 5-point Likert scale. Figure \ref{sus} illustrates the distribution of SUS scores.  The average SUS average score is 89.93, with a standard deviation of approximately 10.01, indicating that participants generally provided high and consistent usability ratings.

Previous research has examined the interpretation of SUS scores in usability assessments, 
highlighting that higher scores are associated with superior usability experiences \cite{bangor2008empirical}. 
Bangor et al. suggest that products with scores above 70 are generally considered usable, 
while those in the high 70s to upper 80s indicate better usability. Truly superior products 
tend to score above 90, reflecting strong user satisfaction. The chatbot's score of 89.93 
places it within this range, suggesting a highly positive user experience.

\subsection{Assessment Regarding Key Content Characteristics} \label{Characteristics}

\begin{figure*}[h]
  \centering
  \includegraphics[width=\linewidth]{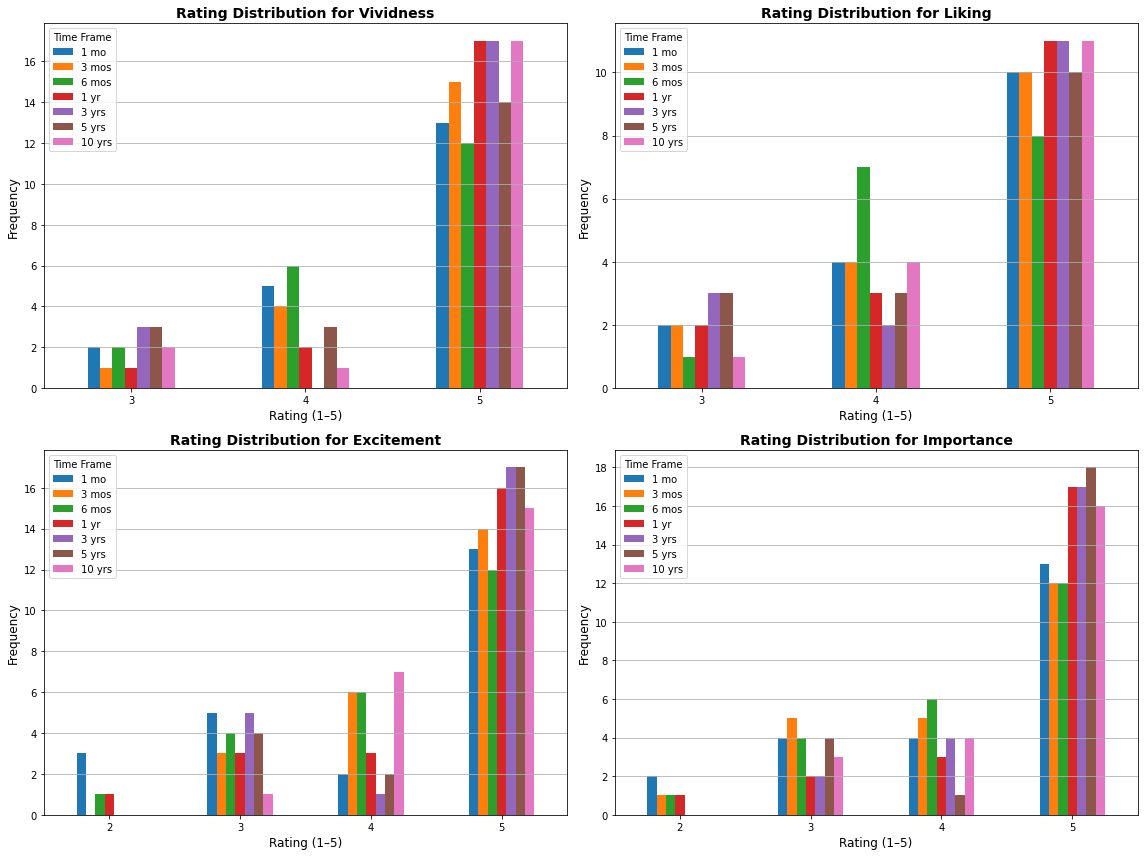} 
  \caption{Ratings for vividness, liking, excitement, and importance characteristics
}
  \label{ratings}
\end{figure*}

To address RQ2 (i.e., What are users’ impressions of the EFT cue texts generated by EFTeacher?), we focused on four key characteristics of the content of the cue texts. 
Figure \ref{ratings} 
displays four frequency distributions (rating values on the x-axis, frequency on the y-axis) for each characteristic, representing ratings over different time frames (1 month, 3 months, 6 months, 1 year, 3 years, 5 years, and 10 years) for the characteristics of Vividness, Liking, Excitement, and Importance. Each plot shows the distribution of ratings on a scale from 1 to 5.

Ratings for Vividness show a strong concentration at the highest rating (5) across all time frames. The frequency bars for rating 5 dominate each time interval, indicating a consistently high perception of vividness. Lower ratings (3 and 4) appear infrequently, supporting the conclusion that participants found the imagined future events vivid.

Ratings for Liking\footnote{How much the participant likes the event} are also strongly skewed toward 5 across time frames. A modest number of responses fall in the 3–4 range, particularly at shorter intervals (e.g., 1 month and 3 months). Nonetheless, since most ratings are 5, it shows that participants generally found the events likable throughout all timeframes.

For Excitement, while rating 5 remains the most frequent, there is a noticeable presence of lower ratings (especially 2, 3, and 4) at earlier time frames, such as 1 and 3 months. As time frames lengthen (e.g., 5 and 10 years), the distribution becomes more concentrated around 5, implying that participants may find longer-term future events more exciting.

Ratings for Importance are generally high, with most bars clustering at rating 5. However, compared to vividness and liking, this category shows more spread in the 2–4 range, particularly in earlier intervals. Their ratings became higher in the later time frames, indicating a growing sense of importance.

The scores of users assigned to cue texts with respect to the four key
content
characteristics
highlight the consistently high median ratings of vividness, liking, excitement, and importance across all timeframes, demonstrating that users perceive EFTeacher as impactful based on what it generated for them through their interactions with it.

\subsection{Assessment Regarding User Perceptions} \label{perception}
To address RQ3 (i.e., 
How do users perceive working with an interactive AI chatbot for EFT generation?),
the research team employed thematic analysis, i.e., a systematic approach to identify themes from user feedback \cite{bowman2023using,coates2021practical,bradley2007qualitative}.
Thematic analysis is a qualitative research approach used to identify and interpret patterns or emerging themes within qualitative data, such as open-ended survey responses or interview transcripts \cite{braun2021thematic}. It can be conducted either deductively, drawing on existing theories and prior research, or inductively, allowing themes to emerge directly from the data itself \cite{braun2021thematic}. Thematic analysis is a process that requires researchers to work with the data through repeated review to develop, refine, and validate the identified emerging themes or patterns. This method has become widely used in HCI research to evaluate user experiences, inform iterative design, and interpret qualitative feedback from system interactions. By applying thematic analysis, researchers can gain rich insights into user experience, which is essential for developing user-centered technologies, including conversational agents.

In our study, we applied this approach to analyze participants’ open-ended feedback.
Three researchers, each with different areas of expertise, independently reviewed the participants’ open-ended responses multiple times. They each developed a list of codes, which are short labels summarizing key points from the feedback. A facilitator, who had expertise in qualitative methods but limited prior knowledge of the survey (to reduce bias), gathered these codes and compiled them into a comprehensive codebook.

The facilitator and coders then engaged in a collaborative session to review and refine the codes. Through some thematic mappings and group discussion, the team merged overlapping codes, clarified distinctions, and identified patterns across responses. This iterative and collaborative process led to the development of broader themes. For each theme, the group defined a clear description and selected illustrative excerpts from participant responses to exemplify its meaning.

Using this finalized codebook, the coders reviewed the user responses again, selecting excerpts that best exemplified each theme. The facilitator aggregated these responses and calculated the reliability between coders, a measure of how consistently the coders applied the codes.
Consistency
was found to be very high, i.e., 94\%. The team then agreed on a final table of themes, including example excerpts to support each one. The analysis provided insights into participants' experiences with EFTeacher, highlighting five distinct themes.

These findings highlighted varying levels of engagement and interaction, offering valuable feedback for understanding how users perceived and interacted with the chatbot.
The next five subsections explain these themes, providing
a framework for understanding participant experiences
and interactions with EFTeacher throughout the study.
To illustrate key points, excerpts from participant responses are included in the text, with participant numbers provided for attribution. As the number of relevant responses is extensive, not all can be discussed in detail here. For a broader view, Table \ref{tab:quotes} presents additional representative quotes across all themes.
\newcommand{\customquote}[1]{\textit{``#1''}}

\begin{table}[H]
\centering
\caption{Example Excerpts for Themes}
\label{tab:quotes}
\renewcommand{\arraystretch}{0.8} 
\begin{tabular}{|p{0.95cm}|p{0.5cm}|p{12.4cm}|}
\hline
\textbf{Theme} & \textbf{ID} & \textbf{Excerpt from Participant Response} \\
\hline
\multirow{7}{*}{1} 
& 112 & “I felt as if I was having a conversation with an instructor!” \\
& 105 &“Actually, the response was better than I would expect a human to respond.” \\

&104& “Even though it was formed from my prompts, it still sounded very scripted"\\
 & 120 & “It was more descriptive than most humans, but with AI, I feel that is what you want.” \\
& 119& “not personal. The repetitiveness made it feel like AI, so not personal."\\
\hline
\multirow{7}{*}{2} 

& 118 & “It gave me good prompts and questions to help me fill in blanks and details about potential future situations.” \\
& 112 & “The program/EFTeacher asked perfect questions to help me envision these events!” \\
& 106 & “It added vivid details that made it much easier to imagine my future.” \\
& 138 & “I liked that the extra details provided a larger picture.” \\
& 115 & “The flowery words used helped see how great the future may be.” \\
\hline
\multirow{4}{*}{3} 
& 118 & “I liked how when I suggested simple edits, it complied easily. ” \\
& 115& “I liked the way it confirmed what you were trying to convey” \\
& 116 & “I enjoy how individual words can be replaced. I did not like an adjective that the boy used and asked for it to replace that specific word” \\

& 120 & “Seems more personable. ChatGPT is overused and not adaptable to the person using it in my opinion.” \\

& 139 & “Very straight forward and easy to change or add details"\\
\hline
\multirow{7}{*}{4 } 
& 137 & “Being prompted to imagine specific details made me reflect on my life” \\
&120& “It was helpful seeing all my events/milestones laid out in a beautiful way"\\
& 106 & “really liked this experience. The redundancy made sense, but it was the only aspect that made it feel slightly less
enjoyable for me. That said, I understand that not everyone envisions their personal journey in the same way or for the
same duration. ” \\
& 125 & “Helped my imagination actually being there. By using more descriptive wording, I felt as if I could really being doing these things and feeling that way"\\
\hline
\multirow{4}{*}{5} 

& 129& “could have a better color scheme; felt like everything kind of blended together with the words at times.” \\
& 137 & “I didn’t like how hard it was to copy what was said. I had a hard time copying the rewrites.” \\
&  139& “larger font sizes would be nice. ” \\
& 117 & “not giving a word count. If there is an expectation of level of detail or goal then nice to be able to track that from the start"\\
\hline

\end{tabular}

\end{table}

\subsubsection{Human-like dialogue and communication proficiency}

Participants reported on EFTeacher’s
dialogue, particularly how conversational, natural, and human-like the interactions felt. A central aspect of
this perception was around the degree to which interactions felt conversational and human-like. Several noted that the system describing its conversational style as authentic and engaging [P105, P109, P112, P118].

They highlighted EFTeacher’s ability to maintain dialogue that resembled human interaction, using descriptors such as natural [P139], genuine [P137], and referencing features like leading questions [P117]. The chatbot’s responses were also praised for being empathetic and contextually relevant:
\customquote{It really interacted with me like a human! I expressed so much joy at something it wrote and said I might need to save it to read later, and it responded so thoughtfully.} P[118]

Despite its strengths, some participants found the interactions  too polished or scripted, detracting from the perceived naturalness P[104, 120, 125], as participant P[115] said:
\customquote{I believe the EFTeacher gave you more than an average human in regards to the descriptive words and paragraphs.}

Repetition also emerged as a common critique. While many users appreciated the flow of the conversation, some noted that EFTeacher tended to repeat similar requests for details, making the interaction feel less dynamic P[119]:
\customquote{Not only was their storytelling great, but it seemed to flow somewhat like a real conversation, except for the repetitiveness of asking for the details it needed each time.} P[123]

These insights suggest that while EFTeacher demonstrates an ability to create engaging and human-like conversations, there is room for improvement.
Participants’ feedback highlights a dual expectation: while many users appreciated EFTeacher's engaging and human-like dialogue, the repetition of similar requests for details caused frustration for some. Striking a balance between maintaining high-quality responses and reducing repetitive patterns is crucial for the refinement of EFTeacher. 

This theme aligns with prior work suggesting that perceived naturalness in chatbot interactions enhances user engagement and the efficacy of conversational agents \cite{li2024finding}. Communication that feels more human-like contributes positively to user participation and
receptiveness to the intervention \cite{olafsson2020motivating}.
Tailoring a chatbot’s persona and communication style to user preferences, and to create a stronger sense of connection, increases the likelihood of sustained engagement \cite{li2023exploring}.

\subsubsection{Adaptability in response to participants’ input}
Participants described how EFTeacher adapted its responses based on their preferences, enabling them to make edits, request word changes, and engage in flexible, personalized interactions. Participants appreciated EFTeacher’s responsiveness to real-time feedback, especially when editing specific words or rephrasing content, and found it easy and intuitive to request changes or content rewrites P[116, 118, 139]:
 \customquote{It was very easy to have EFTeacher make changes. I didn’t know if I could ask for specific words to be changed, so I just tried it, and it worked! When I asked for less vivid wording, it changed it to something that I liked better.} P[112]

Participants noted that EFTeacher not only accepted edits but also  clarified intent by paraphrasing their input or identifying when an input was incomplete P[109]: \customquote{I liked the way it confirmed what you were trying to convey. It was great that EFTeacher understood my words and paraphrased it back to me.} P[115].

They highlighted that EFTeacher felt more grounded and tailored to the user compared to other AI systems P[105, 120].

However, one  participant identified limitations in EFTeacher’s ability to fully simplify responses when
requested. While the tool was adaptable, certain adjustments fell short of expectations:
\customquote{Even when asked to simplify it was still more than I thought a simple description would be. For some people less words work better for their understanding. So when asked to simplify something, I would expect a shorter answer with right-to-the-point statements.} P[125]

Most of the feedback highlights the significance of personalization and adaptability in improving user interaction, emphasizing the importance of addressing diverse user expectations. Prior study \cite{li2024finding} supports this by showing that users highly value the ability to tailor their AI chatbots to their preferences. This customization includes both the chatbot’s visual aspects and its conversational style, reflecting a strong user desire for personalized AI experiences. The present findings underscore the need for adaptive systems when responding to user intent, for user-centered AI systems.

\subsubsection{Imagination facilitated through future-oriented prompts}
EFTeacher supported users in visualizing future events by prompting imaginative thinking. Participants described EFTeacher as a powerful tool for guiding imagination and enhancing their ability to envision detailed, future-oriented scenarios P[127, 123]. They described it as enhancing their milestones and making them feel even more meaningful P[120], painting a vivid picture P[125], and triggering their imagination to visualize these moments P[122]. Many noted helped them articulate life goals and future plans in ways they hadn’t considered before.

The language was often described as descriptive, and the use of descriptive wording transformed abstract thoughts into imagery stories P[122, 115, 104], that participants would not have generated on their own. As one participant shared:
\customquote{Used lots of adjectives. The statement by the AI describes things better than I could put it. It helped me imagine actually being there.} P[125]

Several participants highlighted that the guided prompts and questions posed by EFTeacher were particularly effective in facilitating the imagination of their stories. The questions were described as helpful P[109, 112], well-designed P[103], thought-provoking P[119], and detail-oriented P[137] for storytelling:
\customquote{I liked how it asked questions about my storyline because it made me think about the setting, people, and everything. It really helped bring out the writer in me.} P[122]

EFTeacher also helped participants who struggled to formulate
ideas, particularly for distant future scenarios. The tool’s prompts provided structure and inspiration, helping users
transform ideas into narratives P[127, 123, 106, 137], and through the added details P[106,138]:
 \customquote{I was really struggling to imagine events ten years from now, and it helped me string together some of my piecemeal thoughts!} P[118]

 However, one participant mentioned expecting the chatbot to take the lead in storytelling or in providing meaningful new ideas beyond their initial input:
 \customquote{It felt like I was the one coming up with every story, and they were just riffing on what I came up with by myself. I don’t think this is bad, but I didn’t feel like it was giving me anything to go off of besides the timeframe the story took place in.} P[123]

The ability of EFTeacher to stimulate the imagination through vivid descriptions, questions, and prompts is a standout feature. It has the potential to facilitate imagination, making it an effective tool for users to envision and articulate their future.

\subsubsection{Reflection on using EFTeacher and takeaways from the interaction with EFTeacher}

Participants shared how EFTeacher prompted them to engage in meaningful reflection, fostering emotional insight, future-oriented thinking, and personal awareness. Participants reported that interacting with the system helped them articulate and visualize goals, explore new aspirations, and experience emotional feelings.

Many participants described how EFTeacher encouraged them to reflect on their life direction, consider long-term goals, and think beyond the present moment. The tool provided structured prompts that initiated future-oriented thought P[122, 106, 137, 115]:
\customquote{Don’t typically think very far into the future, and it made me consider where I want my life to be in X months/years.} P[106]

Participants reported that the emotionally rich and vivid content generated by EFTeacher created feelings of excitement P[138] and hope P[107].
Seeing their imagined futures articulated helped some feel encouraged about what lies ahead P[120].

The added details and language
generated with
EFTeacher helped the participants construct emotional narratives P[115, 125] about their future. They noted that the system helped them articulate their thoughts in ways that improved clarity and storytelling:
 \customquote{It was nice that it suggested more helpful details in my writing. It kept me focused and thinking about my future.} P[137]

EFTeacher facilitated personal reflection by encouraging participants to explore future goals and possibilities they had not previously considered. Its vivid and emotionally engaging prompts led users to feel hope, excitement, and validation, helping them reflect on life direction and envision meaningful future scenarios. This aligns with prior research showing how AI tools can support introspection. Similar to AI-driven journaling platforms \cite{nepal2024contextual}, where carefully designed prompts are used to elicit self-reflection and emotional processing, EFTeacher’s personalized and vivid prompts guided users to articulate their aspirations and feelings. Similarly, Liu et al.~\cite{kocielnik2018designing} demonstrate that conversational agents can support reflection in daily life through structured prompts. EFTeacher thus serves not only as an EFT cue generator but also as a tool for self-reflection and future-oriented thinking.

\subsubsection{Feedback about interface, visual design and ease of interaction}

Participants shared a range of impressions about EFTeacher’s interface, regarding its usability and visual design. Overall, many found the interface intuitive and easy to use. For instance, one participant noted, \customquote{It was easy to input information, and the instructions were clear.} P[124]. Others echoed this sentiment, describing the platform as simple to navigate and accessible, especially when submitting responses P[119, 120, 136].

At the same time, participants identified specific areas for improvement that impacted readability, accessibility, and overall user experience. A common concern was the need for better visual clarity, particularly around color contrast P[129] and font P[105].
Others proposed enhancements to make content more accessible, such as audio support P[137] or voice control: \customquote{No option for audio and visual. Hearing the descriptions and scenarios aloud would have added good detail.} P[129].

Usability suggestions focused on improving interaction features that could support task completion. For example, users requested a word count to help guide their responses P[117]. Another participant noted difficulty in copying and reusing texts P[137].

 While the interface was appreciated for its ease of use, incorporating participant feedback could further enhance the user experience across diverse individuals.

\subsection{EFT Cue Text  Analysis}

We analyzed several linguistic features of the EFT cues generated
with
EFTeacher for 25 participants, across seven different timeframes, resulting in a total of 175 cue texts. Table \ref{tab:text-stats} summarizes descriptive statistics for some textual features extracted from cue texts, including:
number of words and unique words, and the Type-Token Ratio (TTR), which quantifies lexical diversity by dividing the number of unique words by the total word count. These metrics were computed individually for each cue and then averaged \cite{lexicalrichness}.
Flesch reading score and Flesch-Kincaid grade level were also computed \cite{textstat}. The Flesch reading score reflects how easy a text is to read, while the Flesch-Kincaid grade estimates the U.S. school grade level required to understand the text. Sentiment polarity and subjectivity were also determined \cite{textblob}. Polarity ranges from -1 (negative) to 1 (positive), and subjectivity from 0 (objective) to 1 (subjective).

\begin{table}[ht]
\centering
\caption{Statistics for Cue Text Features}
\label{tab:text-stats}
\begin{tabular}{|l|r|r|r|r|r|r|}
\hline
\textbf{Feature} & \textbf{Mean} & \textbf{Median} & \textbf{Std Dev} & \textbf{Min} & \textbf{Max} & \textbf{IQR} \\
\hline
Number of words & 108.98 & 106.00 & 25.31 & 41.00 & 220.00 & 31.00 \\
Number of unique words & 76.27 & 75.00 & 15.33 & 35.00 & 135.00 & 18.50 \\
TTR & 0.71 & 0.71 & 0.04 & 0.57 & 0.85 & 0.06 \\
Flesch readability score & 58.74 & 59.03 & 9.44 & 35.01 & 82.48 & 12.75 \\
Flesch-Kincaid grade level & 10.04 & 9.87 & 1.92 & 5.04 & 15.35 & 2.60 \\
Sentiment polarity & 0.33 & 0.32 & 0.14 & -0.08 & 0.82 & 0.19 \\
Sentiment subjectivity & 0.57 & 0.58 & 0.11 & 0.22 & 0.83 & 0.13 \\
\hline
\end{tabular}
\end{table}

According to the table, the average number of words per cue was 108.98, with a median of 106. Cue lengths ranged from 41 to 220 words, indicating a moderately wide variation in text length. The interquartile range (IQR) was 31, suggesting that most cues fall within a relatively consistent length range. 
The average number of unique words per cue was 76.27, with a median of 75 and values ranging from 35 to 135. The standard deviation of 15.33 indicates a moderate degree of variation in vocabulary diversity across cues—some participants used relatively limited vocabularies, while others used a broader range of terms.

The type-token ratio (TTR), a commonly used measure of lexical diversity, was calculated as the ratio of unique words to total words for each cue and then averaged across texts. The mean TTR was 0.71, with a standard deviation of 0.04 and a narrow interquartile range (IQR) of 0.06, suggesting that lexical diversity was relatively stable across cues. A higher TTR indicates a larger proportion of unique words, reflecting more varied language use.

The average Flesch Reading Ease score across cue texts is 58.74, with a median of 59.03, a standard deviation of 9.44, an interquartile range (IQR) of 12.75, and a range from 35.01 to 82.48. The standard deviation and IQR indicate a moderate spread, suggesting that while most cues fall near the average readability level, some are significantly easier or harder to read. A score around 59 falls into the “fairly difficult to read” category \cite{textstat}.
The average Flesch-Kincaid Grade Level is 10.04, with a median of 9.87, a standard deviation of 1.92, and an IQR of 2.60, suggesting that most texts are suitable for readers at the 10th to 12th grade level, with relatively low variability across cues \cite{textstat}.
The average sentiment polarity across cue texts is 0.33, with a median of 0.32, a standard deviation of 0.14, an interquartile range (IQR) of 0.19, and a range from –0.08 to 0.82. The positive mean and median polarity suggest that, on average, cue texts express a positive emotional tone. Importantly, the minimum polarity of –0.08 is very close to zero, indicating that negative sentiment is not observed; even the least positive cues are nearly neutral in tone. Overall the sentiment remains on the positive or neutral side.
The average subjectivity score is 0.57, with a median of 0.58, a standard deviation of 0.11, and an IQR of 0.13, ranging from 0.22 to 0.83. This indicates that the cues are generally moderately subjective, meaning they include personal opinions or affective language rather than purely factual content.

\section{Discussion and Design Recommendation}

The goal of this study is to explore users' experiences and perceptions of EFTeacher, an AI chatbot designed for EFT generation. Specifically, we aim to understand users' overall experiences with the chatbot, their impressions of its functionality, and their perceptions of working with an interactive AI tool for EFT generation.
Prior research has shown that the vividness of EFT cues—measured through self-report \cite{brown2022putting} and content characteristics measurement on a Likert scale \cite{rung2020translating}—is associated with reductions in delay discounting. We capture vividness through participant ratings  
of each cue's vividness, liking, excitement, and importance (Section \ref{Characteristics}),
and through an open-ended survey  (Section \ref{perception}). This allowed us to explore not only whether users experienced cues as vivid, but also when vividness was perceived as helpful (e.g., supporting imagination and motivation) or overwhelming (e.g., excessive verbosity). 
Our thematic analysis indicates that the chatbot's descriptive language was a polarizing feature among participants.
Most
individuals appreciated its vivid descriptions, which enhanced their sense of vividness and made their interactions feel richer. Participants highlighted the chatbot's ability to create detailed visuals and make scenarios feel more tangible. However, others felt that the language was overly elaborate and occasionally unnecessary. For these users, the level of detail made conversations less natural.

Interestingly, this issue could potentially be mitigated through the chatbot's adaptability features, which allow users to request simpler or less descriptive responses. 
Indeed, the chatbot’s adaptability was highlighted as a key strength by some participants, allowing them to refine responses and tailor interactions to their preferences. 
However, some participants did not utilize these features effectively, leading to missed opportunities for improving their experiences. 
This demonstrated a gap in feature accessibility or motivation to adjust responses mid-interaction.
Multiple engagements with the chatbot could facilitate a learning process that helps users become more familiar with these adaptability features, ultimately improving their ability to personalize interactions.

Another notable finding from our thematic analysis is EFTeacher’s ability to facilitate imagination through future-oriented prompts. Participants described how the chatbot’s vivid, descriptive language and structured guidance helped them envision detailed future scenarios, even when initially struggling to generate ideas. Our findings suggest that AI chatbots like EFTeacher can serve not only as tools for cue creation but also as catalysts for future-oriented imagination, helping users simulate possible futures in a structured yet emotionally engaging way.

Participants highlighted the positive emotional impact of their interactions with the chatbot, emphasizing its ability to inspire hope, optimism, and validation. The chatbot’s generated EFTs encouraged users to reflect  on their goals and aspirations, often guiding them to explore aspects of their future they might not have otherwise considered. The vividness and personalization of the chatbot’s EFTs played a key role in creating these positive emotional experiences. EFTeacher demonstrated potential in helping users articulate their thoughts with greater detail and structure, encouraging them to engage more deeply with their personal visions. This approach not only strengthened their emotional connection to the conversation but also nurtured a sense of optimism about the future.
This aligns with the core principles of EFT interventions, where individuals vividly imagine and describe positive future events. This process suggests that an AI chatbot could be a valuable tool for effective EFT interventions by encouraging users to consider the future and helping them focus on their long-term goals.

While the reflective nature of the interactions was highly valued, some participants noted that the level of detail could occasionally feel overwhelming and wordy. This stems from AI-generated language, which, coming from a machine, does not always sound natural. However, despite these constraints, AI can still be a valuable tool for structuring and refining
EFT-related written content, making it a supportive tool,
especially for those who have difficulty with writing.

Participants mentioned that the repetition in the chatbot's responses negatively impacted their interactions. This repetition stemmed from the chatbot prompting users for seven different events at seven different times, following the same process each time. Reducing repetitive patterns could improve the chatbot’s ability to deliver more natural, human-like interactions. To improve user experience, our research group therefore decided to conduct the EFT teacher interview across four (instead of seven) different time frames in future iterations.

Through linguistic feature analysis of cue texts generated by EFTeacher, we found that lexical diversity was high, with a mean Type-Token Ratio (TTR) of 0.71 and an average of 76 unique words per cue. These values indicate a broad range of vocabulary used for the descriptive and detailed language observed in participant feedback. The mean Flesch Reading Ease score was 58.74, classified as “fairly difficult to read” and the average Flesch-Kincaid Grade Level was 10.04, suggesting that the language used may be advanced or verbose for some users. Sentiment analysis revealed a mean polarity of 0.33 (positive) and a subjectivity score of 0.57, indicating that the cues were generally optimistic and included subjective or narrative-style content rather than objective facts.

\subsection{Design Recommendations}
Participants in the study made suggestions that might lead to a next generation version of EFTeacher.
While more development would be needed for such, it is valuable that our study provided a set of design requirements to consider.

\textbf{Integrating Voice Capabilities into EFTeacher}: Based on the findings from our thematic analysis, integrating voice capabilities into EFTeacher could enhance its effectiveness in creating a more engaging and natural conversational experience. Participants  praised EFTeacher's ability to emulate human-like conversation, with its tone being described as thoughtful and lifelike. Introducing voice as a medium of interaction could address any limitations by enhancing the emotional and contextual depth of communication.  Prior research has supported the importance of voice in human-agent interaction. According to \cite{seaborn2021voice}, voice impacts not only the content of the speech but also the overall perception of the agent's personality and context-specific relevance. Additionally, audio delivery has been explored in the context of Episodic Future Thinking (EFT) interventions with promising results. Studies such as \cite{sze2015web, daniel2013future, stein2016unstuck} support the use of audio in EFT interventions. Participants can generate cues and then vividly imagine or listen to audio recordings of their EFT
cues during tasks
that assess
delay discounting. This approach has demonstrated effectiveness in enhancing the intervention experience by leveraging audio's ability to evoke richer emotional and cognitive engagement.

A scoping review of the literature further suggests that voice assistants can serve as a valuable tool in the research on behavioral health interventions \cite{sezgin2020scoping}. This supports the potential for EFTeacher to leverage voice capabilities not only to enhance user engagement but also to support interventions aimed at improving behavioral and emotional well-being.
Voice integration also addresses accessibility concerns raised during usability testing. For example, participants suggested features like voice control to simplify interactions, especially for users who might find typing or reading responses challenging. Voice output and input would not only improve usability but also expand EFTeacher's accessibility to a broader audience.

\textbf{Leveraging Visual Capabilities for EFTeacher}: Advancements in LLMs present a promising avenue for building effective human-AI interaction. The use of multimodal features of LLMs is increasingly gaining attention in recent technologies, as demonstrated in \cite{pereira2024human, lin2024jigsaw, alalyani2024multimodal}. One such feature is the ability of models to generate images from text prompts, as explored in \cite{lin2024jigsaw}. Integrating this capability into EFTeacher could allow the chatbot to produce visual EFTs in response to user queries, making interactions more engaging and particularly useful for tasks that require visual representations, such as illustrating future events.

The benefits of incorporating visual elements are further supported by research on EFT. A study \cite{carr2021written} investigated the effects of drawn and written future cues, and found that both drawn and written future cues led to reduced delay discounting compared to the control group, with drawn cues offering unique advantages in terms of personal relevance and the vivid encoding of details. These findings suggest that enabling EFTeacher to generate or support visual representations of future scenarios could further enhance its impact on future-oriented decision-making.
Moreover, incorporating visual elements could address a key finding from our study: while some participants appreciated the AI-generated language, others found it overly elaborate or unnecessary. This discrepancy highlights the need for alternative modes of interaction, such as visualizations, for users who may not connect with the AI chatbot’s language or who prefer a less verbose experience. By offering visual representations alongside text-based descriptions, EFTeacher could accommodate a broader range of user preferences.

\section{Limitations, Conclusions, and Follow-on Work}
\textbf{Limitations}: 
 Our study design only measured immediate user perceptions and experiences. We did not track longer-term behavioral or psychological outcomes. In subsequent research, the research team aims to conduct longer-term studies to observe how continued use of EFTeacher's generated cues might influence users’ behavior over time. This longitudinal approach would allow us to assess whether the chatbot’s interactive EFT features have a sustained impact on users’ motivation, goal-setting, and health-related behaviors.
 Despite its limitations, our study offers initial evidence supporting the feasibility and promise of a chatbot-based EFT intervention, suitable to guide follow-on studies.
 Our design suggestions could further enhance the application of AI chatbots based on LLMs in EFT interventions for maladaptive health behaviors.

\textbf{Conclusions}: 
In this study, we evaluated EFTeacher, an AI chatbot built on GPT-4-Turbo, designed to assist participants in identifying meaningful future events and in creating detailed text descriptions (cues) for those events. We investigated users’ usability and feasibility experiences with EFTeacher, their impressions of the chatbot, and their perceptions of working with an interactive AI tool for EFT generation.

Our findings highlighted both strengths and areas for improvement in EFTeacher's design and functionality.
Analysis of user impressions, represented through content characteristics of vividness, liking, excitement, and importance ratings, revealed positive trends. 

Analysis of user feedback revealed five key themes that provided deeper insights into users’ experiences with EFTeacher. Participants appreciated the chatbot’s vivid and descriptive language, which enhanced engagement and encouraged self-reflection. This feature helped users articulate their goals and vividly visualize future scenarios. However, some participants found the detailed language excessive, emphasizing the need for adaptability in response style.
The chatbot’s adaptability features—enabling refinements in tone and style—were identified as a significant strength. However, these features were underutilized by some participants, indicating the need for better user guidance. Participants also highlighted challenges with repetitive task structures. Reducing repetitive patterns could improve usability and promote a more natural conversational experience.

Overall, EFTeacher demonstrated potential as a tool for guiding users through EFT generation. Its responses were valued, with participants emphasizing its ability to facilitate imagination and inspire reflection on their goals. The study identified areas for improvement, such as minimizing repetitive patterns and offering clearer guidance on adaptability features. 

Based on feedback from our qualitative research discussed, we refined the chatbot to generate EFT cues for four timeframes (events) to reduce repetition and enhance user experience and engagement, allowing users to discuss events occurring in one month, three months, one year, and three years.

\textbf{Follow-on Work}: 
As a follow-on to the study reported above, we designed a second study, with more participants.
The aims were to: further test our hypotheses, scale up our investigation, and compare EFTeacher with existing alternatives, i.e., to the survey method, or to clinician-led interviews.
Since it was essential to test if EFTeacher led to cues comparable to those from the other methods, and that those cues would reduce delay discounting at least as well as the other methods, we added a final phase to the user session to assess effects on delay discounting.
Since that lengthened the time required of participants, and since we already had feedback and had made changes suggested from open-ended questions, we removed the open-ended questions and reduced the number of generated cues from seven to four.

Preliminary analysis of the second study indicates that EFTeacher helps participants generate cues that are comparable to those of the two other methods.  They also suggest that there is as good an effect on delay discounting as with the other methods, so that at least as good an effect on health outcomes would result from use of EFTeacher.
We plan on subsequent reports of our findings, once we have completed the analysis and comparison of the methods in our second study, the cue texts generated, their reception by participants, and the effect on delay discounting.

\section*{Acknowledgments}
This work was funded by the Seale Innovation Fund and the National Institute of Diabetes and Digestive and Kidney Diseases (5R01DK129567-02S1).


\printbibliography

\end{document}